\documentclass[aps,twocolumn,amsmath,amssymb,showpacs,prb]{revtex4}
\usepackage{epsf}
\usepackage{graphicx}

\newcommand{\etal}{{\it et al.}}

\begin{document}

\title{Momentum anisotropy of the scattering rate in cuprate superconductors}
\author{
        A. Kaminski,$^{1,2,3,4}$
        H. M. Fretwell,$^{3,4}$
        M. R. Norman,$^{2}$
        M. Randeria,$^{5,}$\cite{MR}
        S.~Rosenkranz,$^{2}$
        U. Chatterjee,$^{1,2}$
        J. C. Campuzano,$^{1,2}$
        J. Mesot,$^{6}$
        T. Sato,$^7$ T. Takahashi,$^7$
        T. Terashima,$^8$ M. Takano,$^8$
        K. Kadowaki,$^9$
        Z. Z. Li$^{10}$
       and H. Raffy$^{10}$
      }
\affiliation{
          (1) Department of Physics, University of Illinois at Chicago,
             Chicago, IL 60607\\
          (2) Materials Sciences Division, Argonne National Laboratory,
             Argonne, IL 60439 \\
	(3) Ames Laboratory and Department of Physics and Astronomy,
		Iowa State University, Ames, IA 50011\\
         (4) Department of Physics,University of Wales Swansea,
             Swansea SA2 8PP, UK\\
        (5) Tata Institute of Fundamental Research, Mumbai 400005,
             India\\
         (6) Laboratory for Neutron Scattering, ETH Zurich and PSI
             Villigen, CH-5232 Villigen PSI, Switzerland\\
         (7) Department of Physics, Tohoku University,
             980-8578 Sendai, Japan\\
         (8) Institute for Chemical Research, Kyoto University,
             Uji 611-0011, Japan\\
         (9) Institute of Materials Science, University of Tsukuba,
             Ibaraki 305-3573, Japan\\
         (10) Laboratorie de Physique des Solides,
                  Universite Paris-Sud, 91405 Orsay Cedex, France\\
         }
\date{\today}
\begin{abstract}
We examine the momentum and energy dependence of the scattering rate of 
the high temperature cuprate superconductors using angle resolved photoemission spectroscopy.
The scattering rate is of the form $a + b\omega$ around the Fermi surface for under and optimal doping.  The inelastic coefficient $b$
is found to be isotropic.  The elastic term, $a$, however, is found to be
highly anisotropic for under and optimally doped samples, with an
anisotropy which correlates with that of the pseudogap.  This is contrasted
with heavily overdoped samples, which show an isotropic scattering rate and an absence of the pseudogap above $T_c$. We find this to be a generic property for both single and double layer compounds.
\end{abstract}

\maketitle

\section{Introduction}
There is a general consensus that understanding the normal state excitation spectrum is a prerequisite to solving the high temperature superconductivity problem. Angle resolved photoemission spectroscopy (ARPES) has played an important role in these studies because of the unique momentum and energy resolved information it provides. This includes the observation of a dramatic spectral lineshape change caused by the superconducting transition \cite{NORM97}, the large momentum anisotropy of the superconducting gap consistent with d-wave symmetry \cite{SHEN93,DING96}, an anisotropic pseudogap above $T_c$ \cite{NAT96,LOESER}, the existence of nodal quasiparticles in the superconducting state \cite{ADAM00}, and ``strange metal" behavior above $T_c$ \cite{OLSON90,VALLA99} in the so-called ``normal state". This state is well described by a phenomenological model called the Marginal Fermi Liquid (MFL) \cite{MFL}, however there is as yet no consensus about its microscopic origin. 

A number of studies have been conducted in order to better understand the normal state. Of particular interest is the determination of the scattering rate, which governs the transport properties. 
Earlier on, Valla {\it et al.} showed for optimally doped Bi$_2$Sr$_2$CaCu$_2$O$_{8+\delta}$ (Bi2212) samples that the scattering rate measured at the chemical potential is highly anisotropic around the Fermi surface and it varies linearly with temperature \cite{VALLA00}. This anisotropy was attributed to scattering by out of plane impurities \cite{VARMAIMP,OFF}. In contrast, studies by Bogdanov {\it et al.} \cite{BOGKINK} and Yusof {\it et al.}  \cite{YUSOF} concluded that for overdoped Bi2212 samples, the width of the energy distribution curves (EDCs) (related to the scattering rate) at the chemical potential is isotropic around the Fermi surface. 
Although these studies of the scattering rate around the Fermi surface provide useful information, they are limited in their scope. They measure or infer the scattering rate ({$\it Im\Sigma$}) at a single energy ($\omega$) and do not extract its functional form. Other studies that investigate the scattering rate as a function of energy are limited to the nodal direction alone \cite{ADAM01,BOGKINK, YUSOF}. 

The data presented in this paper provide a comprehensive measurement of the functional form of the scattering rate as a function of energy and momenta around the Fermi surface. Consequently we are able to shed light on the nature of the excitations. We have conducted measurements of the scattering rate in under, optimally and overdoped samples and examine how it relates to the anisotropy. 
The key findings are:  1)  In under and optimally doped samples, which show an anisotropic scattering rate around the Fermi surface, there is a linear dependence of the scattering rate as a function of energy (for all points around the Fermi surface); 2) In contrast, for overdoped samples there is no anisotropy around
the Fermi surface and the scattering rate is more consistent with a superlinear behavior with energy; 
3) We also find that the presence of the scattering rate anisotropy in under and optimally doped samples and its absence in overdoped samples applies equally well to both single and double layered compounds, which suggests that bilayer splitting is not responsible for the anisotropy;  
4) For under and optimally doped samples, the momentum dependence of the pseudogap mirrors the momentum dependence of the scattering rate, which implies that both may be caused by the same interaction;
5) The functional form of the scattering rate for under and optimally doped sample is $a+b\omega$, where the inelastic parameter, $b$, is isotropic around the Fermi surface; and finally, 6) We show that the bare Fermi velocity can be directly obtained from ARPES. 

\section{Experimental details}
Most samples employed for this work are single crystals grown using the floating zone method. They include optimally doped Bi2212 samples ($T_c$=90K) (used in an earlier study \cite{ADAM01}), as well as heavily overdoped ($T_c \sim 0$) Bi$_{1.80}$Pb$_{0.38}$Sr$_{2.01}$CuO$_{6-\delta}$ (Bi2201) \cite{SATO01}.
The overdoped Bi2212 and optimally doped Bi$_{2}$Sr$_{1.6}$La$_{0.4}$CuO$_{y}$ thin film samples (the latter referred to as optimally doped Bi2201) were grown using an RF sputtering technique. The samples were mounted with $\Gamma - M$ parallel to the photon polarization \cite{FOOT1} and cleaved in situ at pressures less than 2$\cdot$10$^{-11}$ Torr. Measurements were performed at the Synchrotron Radiation Center in Madison Wisconsin, on the U1 undulator beamline supplying $10^{12}$ photons/sec, using a Scienta SES 200 electron analyzer with an energy resolution of 16 meV and momentum resolution of 0.01 $\AA^{-1}$ for a photon energy of 22 eV.

\begin{figure}
\includegraphics[width=3.4in]{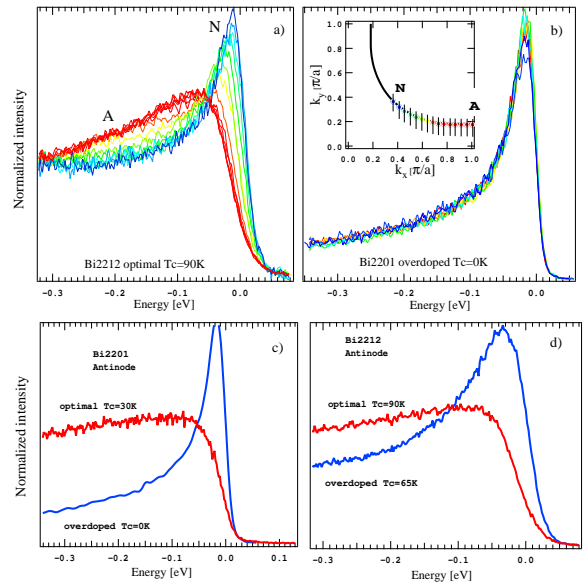}
\caption{Energy distribution curves (EDCs) along the Fermi surface: 
a) EDC data from optimally doped Bi2212 ($T_c$=90K) at T=140K. 
b) the same for overdoped Bi2201 ($T_c \sim 0$) at T=40K.  
 The inset shows the color coded points on the Fermi surface where the EDCs shown in a) and b) were measured (N is the node and A is the antinode of the d-wave gap).
c) Comparison of the EDCs at the antinode for optimally and overdoped Bi2201 at T=40K. d) the same (at the bonding Fermi surface) for optimally and overdoped Bi2212 at T=100K.
}
\label{fig1}
\end{figure}

\section{Scattering rates and renormalized Fermi velocity}
In Figs.~1a and b, we plot energy distribution curves (EDCs) along the Fermi surface in the pseudogap state of optimally doped Bi2212 and the normal state of highly overdoped Bi2201, respectively. These data reveal that in the optimally doped case, there is a strongly anisotropic pseudogap which is zero in an arc centered at the node of the d-wave superconducting gap ($\Gamma-Y$ Fermi crossing), and takes its maximal value at the antinode ($M-Y$ Fermi crossing).
Moreover, there appears to be a strong anisotropy of the scattering rate, since the spectral peaks at the antinode are much broader than at the node.  Although this has been suggested to be due to an unresolved energy splitting caused by bilayer mixing \cite{BILAYER}, 
 in Fig.~1c, we show data at the antinode for optimally doped single layer Bi2201, which has similar spectral characteristics to that of Bi2212 (shown in panel d), arguing against a bilayer effect as the origin of the anisotropy.
We can contrast the behavior shown in Fig.~1a with that of heavily overdoped Bi2201 in the normal state, where no energy gap is present (Fig.~1b).  In this case,
the spectral peak is isotropic around the Fermi surface, indicating that the scattering rate is also isotropic. A similar conclusion was
reached in recent studies of heavily overdoped Bi2212 samples where strong bilayer splitting is present \cite{BOG02,YUSOF}. 

\begin{figure}
\includegraphics[width=3.4in]{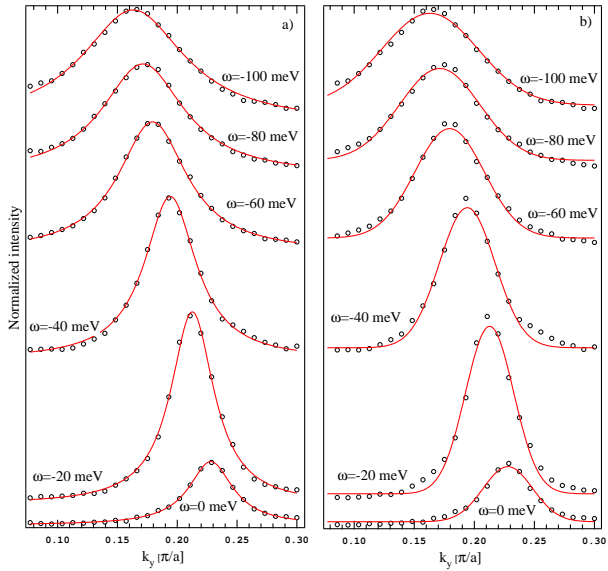}
\caption{Quality of the fitting to the Momentum Distribution Curves (MDCs). Data for an optimally doped Bi2212 sample, same as in  Fig. 1a. Each curve corresponds to a different energy ($\omega$) and is measured along the $k_y$ direction at $k_x$=0.59:
a) lorentzian fits to the data,
b) gaussian fits to the data.
}
\label{fig2}
\end{figure}

To obtain more quantitative information, we analyze momentum distribution curves (MDCs) \cite{VALLA99}.  As we pointed out earlier \cite{ADAM01}, the MDC halfwidth (in the absence of an energy gap) is equal to the imaginary part of the self-energy at that energy, $Im\Sigma(\omega)$, divided by the bare Fermi velocity, $v_{F0}$ (not the renormalized one, $v_F$).  Also, the peak position of the MDC (as a function of binding energy) gives the band dispersion. 

First we examine the quality of the fits to the MDCs. Figure 2 shows sample data from Bi2212 with lorentzian (panel a) and gaussian (panel b) fits. The lorentzian fits are a much better representation of the data at all energies, with a chi-squared statistic several times lower than for the gaussian fits.

\begin{figure}
\includegraphics[width=3.4in]{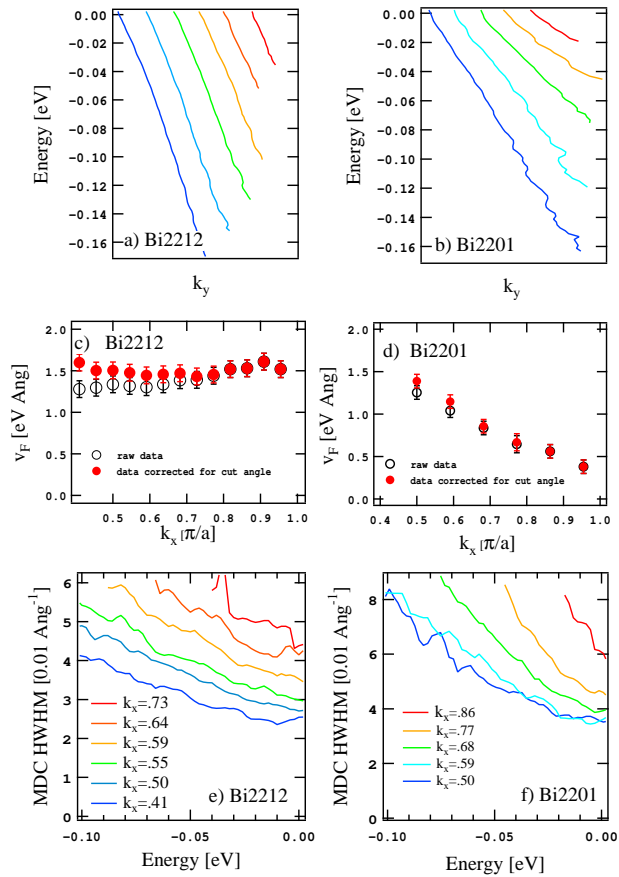}
\caption{The data obtained from MDCs (same as in Fig.~1a and 1b) for selected momentum cuts parallel to $M-Y$ (labeled by the $k_x$ value along $\Gamma-M$, see inset to Fig.~1b). (a,b) Band dispersion. (c,d) Fermi velocity - raw data are shown as empty circles, the data corrected by the angle of the cut are shown as dots. (e,f) peak widths. 
a), c), e) optimally doped Bi2212 at T=140K and b), d), f) overdoped Bi2201 at T=40K.
}
\label{fig3}
\end{figure}

In Fig.~3, we plot the dispersion, Fermi velocity and peak widths obtained from the MDCs along selected cuts in momentum space, parallel to the $M-Y$ direction. 
The first interesting point to note is that the slopes of the MDC dispersion curves are fairly similar in optimally doped Bi2212 (panel a) and vary much more in overdoped Bi2201 (panel b). The slope of the MDC dispersion at the chemical potential is equal to the renormalized Fermi velocity. We extract this from the MDC dispersion and plot it in panels (c) and (d). The Fermi velocity for optimally doped Bi2212 (panel c) is roughly isotropic, a conclusion also reached in an earlier ARPES study \cite{VALLA00}. 
Since in the nodal region the cuts along which the data were obtained are not perpendicular to the Fermi surface, we also show the velocities that were corrected by taking into account the angle of the cut  
versus the Fermi surface normal (solid circles).  
We now examine the renormalized Fermi velocity in heavily overdoped Bi2201 (panel d). In contrast to optimally doped Bi2212, this is highly anisotropic and is much smaller at the antinode; a result consistent with a previous tight binding fit to normal state ARPES dispersions in overdoped Bi2212 \cite{NORM95}. The strong anisotropy observed in Bi2201 arises from the close proximity of the saddlepoint of the band dispersion at $(\pi,0)$ (where the velocity is zero) to the Fermi energy for this heavily overdoped sample \cite{SATO01}. Stated another way, the small anisotropy in the optimally doped case implies that the saddlepoint in the band structure at $(\pi,0)$ is significantly far away from the Fermi energy.

This effect is illustrated in Fig.~4a. Here we plot the band dispersion obtained from a tight binding fit 
where the saddlepoint has been assumed to lie far below (150 meV) the chemical potential - the same energy as at $(\pi,0)$ in optimally doped Bi2212. The band dispersion from the tight binding fit is plotted along cuts taken in the same geometry as the experimental data (Fig.~3). In panel (b) we show the
Fermi velocity calculated from the slope of the band dispersion. This velocity bears a close resemblance to the real data for optimally doped Bi2212 in the sense that is roughly isotropic (Fig.~3c). 
We now address panels (c) and (d), which show the tight binding dispersion and Fermi velocity, respectively, in the case where the saddlepoint at $(\pi,0)$ lies near (10 meV below) the chemical potential - similar to the situation in overdoped Bi2201\cite{SATO01}. The effect of the saddlepoint is readily apparent as the velocity decreases quite rapidly near the $(\pi,0)$ point - exactly the same as in the experimental data (Fig.~3d). 
In summary, our results for the Fermi velocity are consistent with tight binding calculations and suggest that renormalization effects in the normal state are momentum independent. This strongly supports our conclusion (see later) that the $b$ term of the self energy in under and optimally doped Bi2212 is also momentum independent (isotropic).
 
 \begin{figure}
\includegraphics[width=3.5in]{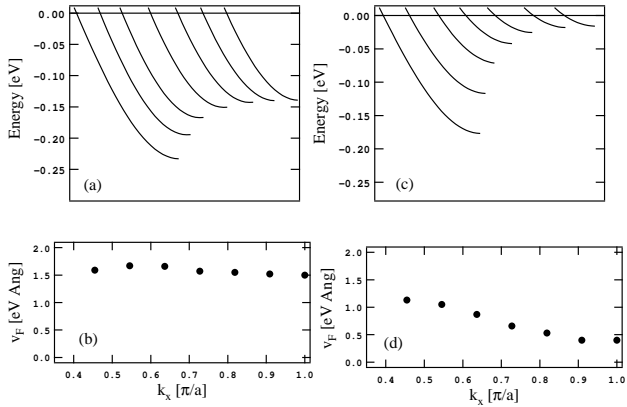}
\caption{The band dispersion (a,c) and renormalized Fermi velocity (b,d) obtained via a tight binding model where the saddlepoint at $(\pi,0)$ lies at (a,b) 150 meV below the chemical potential (equivalent to optimally doped Bi2212) and (c,d) 10 meV below (equivalent to overdoped Bi2201). 
}
\label{fig4}
\end{figure}

\section{Functional form of the self-energy}
The energy dependence of the MDC peak widths for the various momentum cuts is shown in Figs.~3e and 3f.  
Consider first the optimally doped Bi2212 data (3e). To a good approximation, the result for any
 particular cut can be fit to the form $a_M + b_M\omega$ \cite{FOOT5}.
This is analogous to the $a_M + b_MT$ form indicated for the temperature dependence of MDC widths previously reported by Valla \etal \cite{VALLA00}.   
 
\begin{figure}
\includegraphics[width=3.5in]{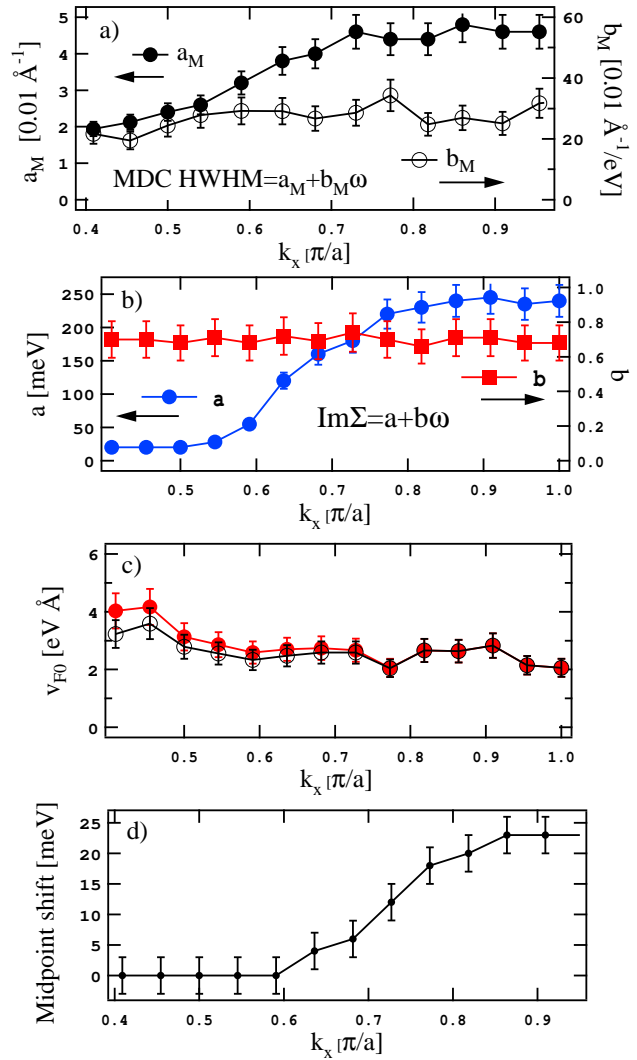}
\caption{The pseudogap and the elastic (``$a$'') and inelastic (``$b$'') portions of the scattering rate ($Im\Sigma$=$a+b\omega$) around the Fermi surface for optimally doped Bi2212.
a) momentum dependence of the $a_M$ and $b_M$ terms obtained by fitting MDC HWHM data from Fig.~3e.
b) momentum dependence of the $a$ and $b$ terms obtained by fitting EDCs from Fig.~1a. 
c) bare velocity obtained by dividing the $b$ and $b_M$ coefficients - raw data are shown as empty circles, the data corrected by the angle of the cut are shown as dots.
d) Position of the midpoint of the leading edge of the EDC around the Fermi surface for optimally doped Bi2212 obtained from Fig.~1a. This is
an approximate measure of the pseudogap - the actual value of the pseudogap is equal to about twice the midpoint shift \cite{PHENOMENOLOGY}. ($k_x$ labels the momentum cut as in Fig.~3, with $k_x=0.4$ corresponding to the node and $k_x=1.0$ to the antinode). 
}
\label{fig5}
\end{figure}

 In Fig.~5a, we show the momentum dependence of the $a_M$ and $b_M$ terms extracted from the MDC HWHM for optimally doped Bi2212 in Fig.~3e.  To complement these results, we have also fit the EDCs along the Fermi surface for this sample using a model self-energy, $\Sigma$.  We have tested both quadratic and linear energy dependences for $Im\Sigma$ and found that only the latter gives an adequate description of the data.  $Re\Sigma$ is determined by Kramers-Kronig transformation of $Im\Sigma=a+b\omega$, assuming the latter saturates at a constant value beyond a cutoff energy of 0.5 eV.  The background signal was determined using methods described previously \cite{BACKGROUND} and added to the calculated curves. The experimental energy resolution was taken into account by convoluting the calculated curves with the appropriate gaussian function. The experimental momentum resolution was taken into account by summation of the spectra over the analyzer momentum window. The quality of the fits is illustrated in Fig.~6 along with plots of the relevant self energies $\Sigma$. To improve the determination of the $b$ coefficient we have performed fits to EDCs peaked at high binding energy, where $b$ has the biggest impact on the lineshape. The EDC fits performed close to the Fermi momentum allow a precise determination of the $a$ coefficient. In Fig.~5b, we show the values of the $a$ and $b$ coefficients obtained from these EDC fits \cite{FOOT2}. Note the similarity of the results to those in panel a) despite the quite different methodologies used. Hence, we can have confidence in the validity of our results for the $a$ and $b$ parameters.
 
\section{Extraction of the bare Fermi velocity from ARPES} 
As an interesting aside, we can exploit the above to estimate in a very simple way the bare velocity.
We have previously shown\cite{ADAM01} that the width of the MDC peak is given by $W_M = \Sigma^{\prime\prime}(\omega)/v_{F0}$. Since now $W_M =a_M+b_M \omega$ and $Im\Sigma$=$a+b\omega$, it follows that  $v_{F0}=b/b_M$, therefore, we can extract the bare velocity around the Fermi surface directly from our data as shown in Fig.~5c\cite{FOOTB}. We note that the bare velocity obtained in this simple way is consistent with band theory predictions and is in agreement with  alternative method of extracting bare velocity\cite{BORISENKOVF}.
 
\begin{figure}
\includegraphics[width=3.5in]{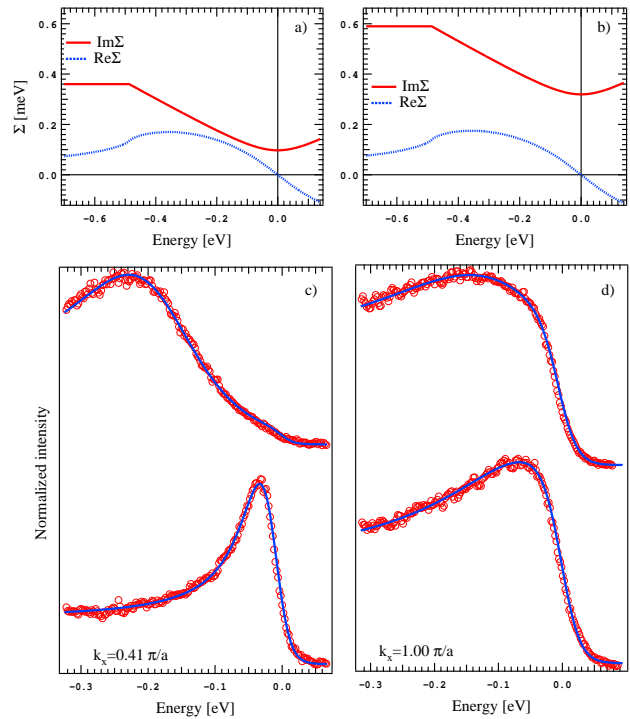}
\caption{Fits to energy distribution curves using an a+b$\omega$ model for the self energy.
a) real and imaginary parts of the self energy used to fit the EDC data in the nodal region.
b) real and imaginary parts of the self energy used to fit the EDC data in the antinodal region.
c) EDC data and fits in the nodal region. The upper curves are close to the band minimum ($k_y$=0), while the lower curves are close to the Fermi momentum.
d) EDC data and fits in the antinodal region. The upper curves are close to the band minimum ($k_y$=0), while the lower curves are close to the Fermi momentum.
}
\label{fig6}
\end{figure}

\section{Conclusions}
Returning to Fig.~5a and 5b, the first conclusion to draw from the data is that the $b$ ($b_M$) term is isotropic in both plots. 
At first sight, it would appear that the Bi2201 case is different, since the $b_M$ term in
that case (slope of the curves in Fig.~3f) appears to increase as the antinode is
approached.  But once the velocity is divided out (Fig.~3d) \cite{FOOT3}, we find in this case as
well that the $b$ term for $Im\Sigma$ is isotropic, which is consistent with
the isotropy of the EDC lineshapes shown in Fig.~1b.
The isotropy of $b$ provides strong support of the original marginal Fermi liquid conjecture \cite{MFL}, where isotropic (i.e., local) behavior is required to guarantee~$\omega/T$ scaling.

The $a_M$ term (obtained from the zero intercept in Fig.~3e) is highly anisotropic. This is consistent with the strong anisotropy of the EDC lineshapes shown in Fig.~1a.
The anisotropy of the $a$ term in optimally doped samples has been attributed to off planar impurities \cite{VARMAIMP,OFF}. On the other hand, we note the remarkable similarity between the anisotropy of this term and that of the pseudogap\cite{FOOT4} (Fig.~5d).  This indicates to us that the anisotropy is probably not due to impurity scattering, but rather is related to the same interaction that gives rise to the pseudogap. This is consistent with the observation of isotropic lineshapes for more heavily overdoped samples of Bi2212 where no pseudogap is present. However, other possibilities could also be considered, such as the cold spots model of Ioffe and Millis, where a highly anisotropic scattering rate is conjectured due to scattering from d-wave pairing fluctuations \cite{COLDSPOT,HUSSEY}.   

Finally, we note that an anisotropy in the ``zero intercept" of the MDC width is also evident in the heavily overdoped Bi2201 sample (Fig.~3f). But, in this case, it can be accounted for by the anisotropy in the bare Fermi velocity, which decreases towards the antinode. Thus the $a$ term in $Im\Sigma$ is isotropic (again, this is consistent with the isotropy of the EDC lineshapes shown in Fig.~1b). 

In conclusion, we find that the normal state scattering rate in under and optimally doped cuprates can be approximated by the form $a + b\omega$.  The inelastic $b$ term is found to be isotropic, which is a necessary ingredient in the marginal Fermi liquid conjecture \cite{MFL}.  In contrast, the $a$ term is found to be anisotropic for under and optimally doped samples, with the anisotropy linked to that of the pseudogap.  

\section{Acknowledgements}
This work was supported by the NSF DMR 9974401, the U.S. DOE, Office of Science, under Contract No. W-31-109-ENG-38 and the MEXT of Japan. The Synchrotron Radiation Center is supported by NSF DMR 9212658.  Ames Laboratory is operated for the U. S. DOE by Iowa State University under contract No. W-7405-Eng-82. AK was supported by the Royal Society of Great Britain, 
and MR by the Indian DST through the Swarnajayanti scheme.  We acknowledge useful discussions with E. Abrahams, C.M. Varma, and P.D. Johnson.


\begin{thebibliography}{99}
\bibitem{FOOT1}
Zone notation is $\Gamma$ $(0,0)$, $M$ $(\pi,0)$, and $Y$ $(\pi,\pi)$ in units
of $1/a$, where $a$ is the Cu-Cu distance.
\bibitem{FOOT5}
Note that close to $\omega=0$ the MDC peak width ``levels off" due to the temperature term in  Im$\Sigma$ and the influence of experimental resolution.
\bibitem{FOOT2}
In these EDC fits, the pseudogap was taken into account in the simplest
possible manner by displacing the spectral peak to a larger binding energy.
\bibitem{FOOTB}
We use the $b$ term to estimate the bare velocity, as it is less subject to resolution and
interpretation issues than the $a$ term (see Ref.~\onlinecite{FOOT4}).
\bibitem{FOOT3}
In the heavily overdoped case, where the renormalizations are weak and isotropic,
we can make the approximation of using the true Fermi velocity rather than the
bare one.
\bibitem{FOOT4}
One may here consider the MDCs obtained from a broadened version of BCS theory \cite{MDC01}.  In this case, the self-energy is of the form $i\Gamma + \Delta^2/(\omega+\epsilon_k-i\Gamma)$
where $\Gamma$ is the lifetime broadening, $\Delta$ the energy gap, and
$\epsilon_k$ the normal state dispersion.  At the Fermi energy ($\omega=0)$,
the resulting MDC HWHM is found to be equal to $\sqrt{\Gamma^2+\Delta^2}/v_{F0}$.
In the limit of small $\Gamma$, the anisotropy of the MDC HWHM would follow that of $\Delta$. This argument, though, would not explain the anisotropy of the EDC lineshapes at
high binding energies observed in Fig.~1a.
\end{thebibliography}
\end{document}